\documentclass{PoS}

\newcommand{\ba}{\begin{array}{c}}
\newcommand{\ea}{\end{array}}

\newcommand{\be}{\begin{equation}}
\newcommand{\ee}{\end{equation}}

\newcommand{\ket}{\,\rangle}
\newcommand{\bra}{\langle \,}

\newcommand{\mM}{\mathcal{M}}

\newcommand{\Frac}[2]{\frac{\displaystyle #1}{\displaystyle #2}}

\title{Chiral Dynamics Predictions for $\eta'\rightarrow\eta\pi\pi$}

\ShortTitle{$\eta'\rightarrow\eta\pi\pi$}

\author{\speaker{Pere Masjuan}\thanks{This contribution is based on preliminary results and on work in progress ~\cite{RJJP}. I would like to thank the organizers for the nice atmosphere during the conference. This work has been supported by CICYT-FEDER-FPA2008-01430, SGR2005-00916, the Spanish Consolider-Ingenio 2010 Programme CPAN (CSD2007-00042) and by the EU Contract No. MRTN-CT-2006-035482, FLAVIAnet.}\\
        Grup de F\'{\i}sica Te\`{o}rica \& IFAE, Universitat Aut\`{o}noma de Barcelona, E-08193 Bellaterra, Catalunya, Spain.\\
        Fakultät für Physik, Universität Wien, Boltzmanngasse 5, A-1090 Wien, Austria.\\
        E-mail: \email{masjuan@ifae.es}}

\abstract{We study the decay $\eta'\rightarrow\eta\pi\pi$ in two different chiral invariant approaches: Large-$N_c$ Chiral Perturbation Theory (ChPT) and Large-$N_c$ Resonance Chiral Theory (RChT). We analyze the Dalitz plot and the invariant mass spectra. We also compare the relevance of the isoscalar and isovector channels in these approaches. While the predictions of Large-$N_c$ ChPT at next-to-leading order slightly disagree with the measured decay width (showing the need for final state interactions and higher order local contributions), a reasonable agreement is obtained for the case of RChT. Forthcoming experimental analyses at Bonn, Frascati, J\"ulich and Mainz will decide among the different frameworks.}

\FullConference{6th International Workshop on Chiral Dynamics, CD09\\
		July 6-10, 2009\\
		Bern, Switzerland}

\begin{document}

\section{Introduction}


The decay modes of the $\eta$ and $\eta'$ mesons are special laboratories for studying symmetries and symmetry breakings in QCD. Specially, they can help us to test our Effective Field Theories in a chiral invariant framework. Even more since different upgrades are taking place in several accelerator research infrastructures such as Crystal Ball at MAMI-C, Crystal Barrel at ELSA, KLOE2 at DAPHNE and WASA at COSY. With all these expected data, one can determine the Dalitz plot parameters and then compare with the EFT predictions to test their accuracy.


In particular, $\eta'$ decays are dominated by strong interactions since almost $65\%$ of them are done through $\eta'\rightarrow\eta\pi^+\pi^-$ and $\eta'\rightarrow\eta\pi^0\pi^0$ ($44.6\pm1.4\times 10^{-2}$ and $20.7\pm0.9\times 10^{-2}$ \cite{PDG09}, respectively). For that purpose, we will use the most general Lagrangian according to chiral symmetry and also the resonance Lagrangian, both in the Large-$N_c$ limit, to show that these $\eta'$ decays are clean processes to study scalar mesons and some of their properties.

With these tools in hand we present the predictions for the Dalitz plots and the invariant mass spectra in both Large-$N_c$ ChPT (section 2) and Large-$N_c$ RChT (section 3). All the details, discussions and further comparisons are still preliminary and will be found in Ref.~\cite{RJJP}.

\section{Large-$N_c$ ChPT}

From a general point of view, the theoretical tool to study $\eta$ decays ($m_{\eta}\simeq 550$ MeV) is Chiral Perturbation Theory (ChPT) \cite{GL85}. It is an effective field theory based on the Goldstone Bosons as degrees of freedom from the spontaneous breaking of the chiral symmetry. It is an expansion in momenta and in quark masses and its breakdown scale is at the scale of the firsts resonances, which are not included in the description. This happens to be at the scale of the $\rho$-meson mass, depending on the particular channel.
In order to treat the $\eta'$ in this framework, the calculation needs to be organized in a slightly different way, because $m_{\eta'}\simeq 960$ MeV, and then $SU(3)$-ChPT can not deal with it. However, the mechanism that gives mass to the $\eta'$, the $U(1)_A$ anomaly, it turns to be suppressed if the number of colors $N_c$ is not kept at three but sent to infinity \cite{tHooftLN}. Actually, also other aspects of QCD simplify. In this framework, the $\eta'$ becomes then a Goldstone Boson and can be introduced in the Lagrangian. This new degree of freedom in the Lagrangian implies the enlarged $U(3)_L\times U(3)_R$ global symmetry \cite{LKlargeN}.

The leading order prediction from this lagrangian to the decay $\eta'\rightarrow\eta\pi\pi$ happens to be specially suppressed, being proportional to $m_{\pi}^2$ and giving a prediction too much off the experimental data \cite{BijnensLO}. However, going one step further, the next-to-leading order (NLO) calculation in the large-$N_c$ ChPT expansion in $p^2,\,m_P^2$ and $1/N_c$ \cite{LKlargeN} produces a drastic improvement of the decay rate, with the right order of magnitude of the process. This means that the dominant piece of this decay arises at NLO. Indeed, in the massless quark limit the LO term vanishes.

From the Large-$N_c$ perspective, it is very convenient to express the amplitude in terms of OZI-allowed and OZI-suppressed components:
\vspace{-0.3cm}
\begin{eqnarray}
\mM_{\eta'\to\eta \pi\pi} &=&
c_{qq} \, \mM_{\eta_q \eta_q \pi\pi}
   +
c_{sq} \, \mM_{\eta_s \eta_q \pi\pi}
  +
c_{ss} \, \mM_{\eta_s \eta_s \pi\pi}\, .
\end{eqnarray}

\noindent
The factors $c_{qq}$, $c_{sq}$, $c_{ss}$ are given by the $\eta-\eta'$ mixing \cite{RJJP} with the numerical values $F_1=1.1F_{\pi}$, $F_8=1.3F_{\pi}$, $\theta_1=-5^\circ$ and $\theta_8=-20^\circ$ \cite{LKdecays}. On the other hand, the pieces $\mM_{\eta_a\eta_b \pi\pi}$ are independent of the $\eta-\eta'$ mixing and define the dynamics of the system. The kinematics are given in terms of $s,t,u$ defined by:
\vspace{-1.0cm}
\begin{center}
\begin{eqnarray}
s\,=\,(p_{\pi^+}+p_{\pi^-})^2\,=\,(p_{\eta'}-p_{\eta})^2 & \quad t\,=\,(p_{\pi^+}+p_{\eta})^2\,=\,(p_{\eta'}-p_{\pi^+})^2\\\nonumber
u\,=\,(p_{\pi^-}+p_{\eta})^2\,=\,(p_{\eta'}-p_{\pi^-})^2 & s+t+u\,=\,m_{\eta'}^2+m_{\eta}^2+2m_{\pi}^2 \, .
\end{eqnarray}
\end{center}

In the Large-$N_c$ ChPT at NLO, the amplitude reads:
\vspace{-0.1cm}
\begin{eqnarray}
& & \mM_{\eta'\to\eta\pi^+\pi^-} = \,c_{qq} \times  {1 \over  F^2}\Bigg[ \frac{m^2_{\pi}}{2} - \frac{2 L_5 m^2_{\pi}}{F^2}\bigg( m^2_{\eta'}+m^2_{\eta}+2 m^2_{\pi} \bigg)+\\
& + & \frac{2(3L_2+ L_3)}{F^2}\bigg(s^2+t^2+u^2-(m^4_{\eta'}+m^4_{\eta}+2 m^4_{\pi})  \bigg)+ \frac{24 L_8 m^4_{\pi}}{F^2}+ \frac{2 \Lambda_2  m^2_{\pi}}{3} \Bigg] \, + \, c_{sq} \times \frac{\sqrt{2} \Lambda_2 m^2_{\pi}}{3 F^2},\nonumber
\end{eqnarray}

\noindent
where the piece $\mM_{\eta_s\to\eta_s\pi^+\pi^-}=0$ (up to the considered order). In the isospin limit, assumed here, the neutral pion decay is related to this through $\mathfrak{B}(\eta'\to\eta\pi^+\pi^-)=2 \mathfrak{B}(\eta'\to\eta\pi^0\pi^0)$. The largest contribution comes from the $3L_2+L_3$ term, which is proportional to the external momenta. Everything else is proportional to different powers of $m_{\pi}^2$ and, then, suppressed. Using the leading order contribution and the dominant piece with $3L_2+L_3=(3\times1.8-4.3)10^{-3}$ \cite{PichLN}, one obtains a prediction for the branching ration of $\mathfrak{B}(\eta'\to\eta\pi^+\pi^-)=48.6\%$. Nevertheless, the interference term in the squared amplitude $|\mM_{\eta'\to\eta\pi\pi}|^2$ is not negligible numerically and then the $m_{\pi}^2$ terms must be kept, especially that concerning $L_5$. The final prediction for the referred inputs is $\mathfrak{B}(\eta'\to\eta\pi^+\pi^-)=26.3\%$.

\begin{figure}
\centering
  \includegraphics[width=2in]{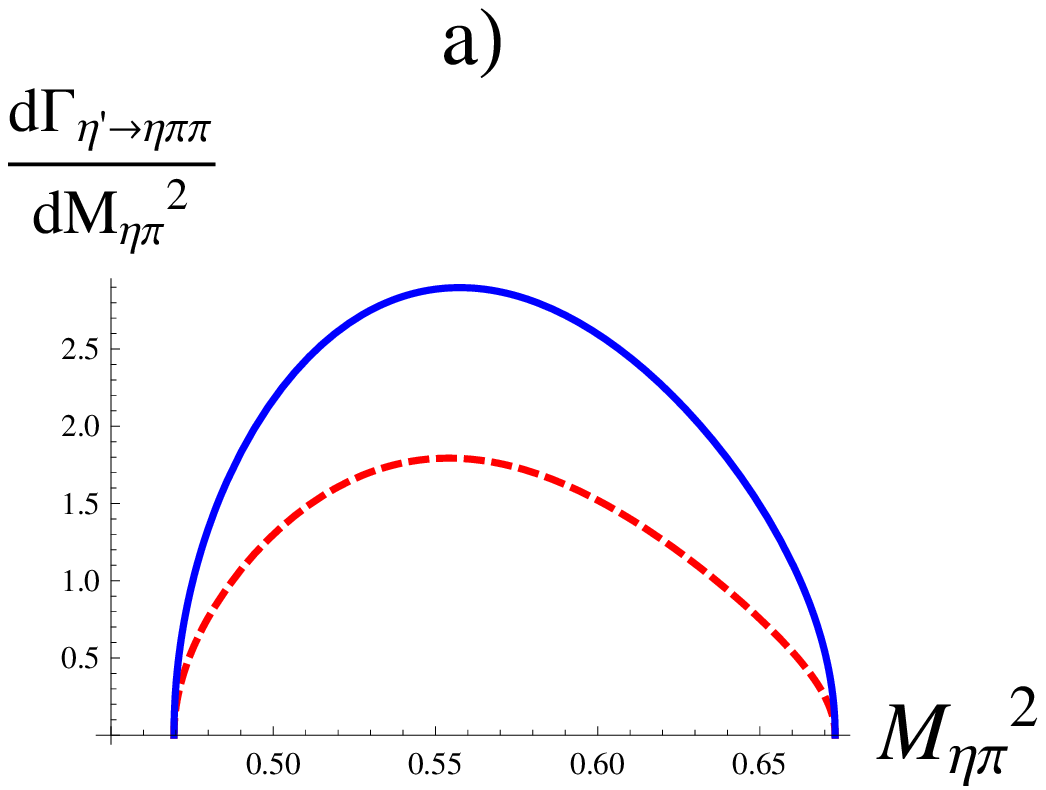}
  \includegraphics[width=2in]{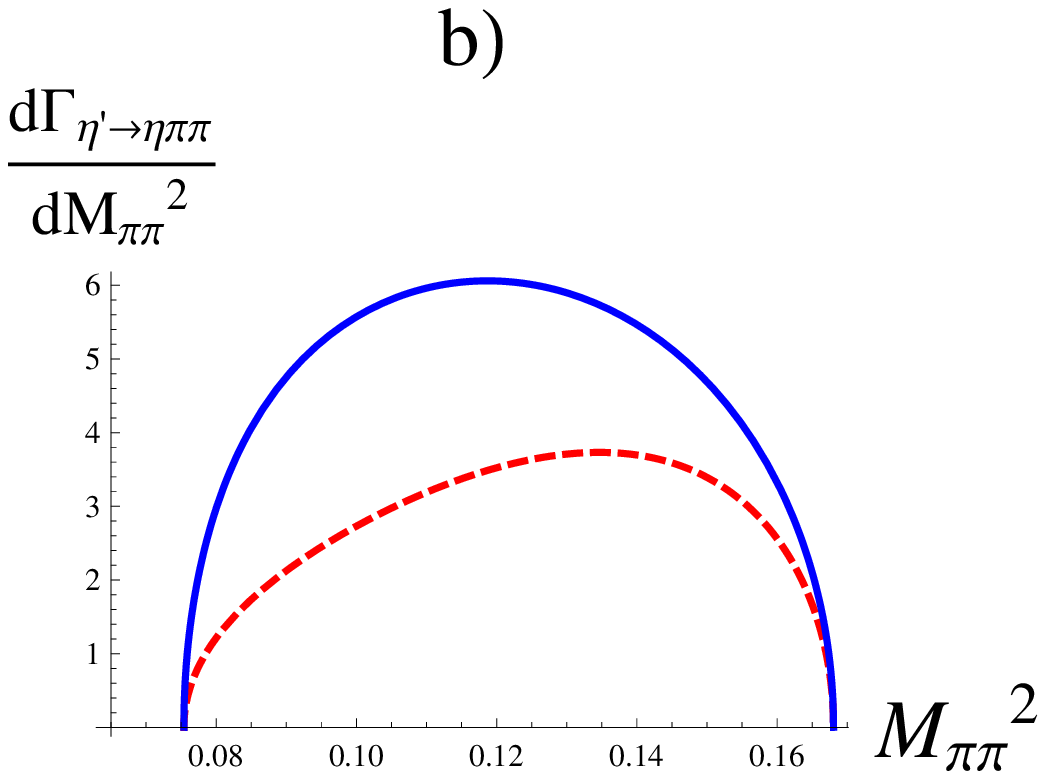}\\
  \caption{Invariant mass spectra predictions for the Large-$N_c$ ChPT (dashed-red) and for the Large-$N_c$ RChT (solid-blue) in the $t$-channel (a) and in the $s$-channel (b).}\label{LNRspect}
\end{figure}

\begin{figure}
\centering
  \includegraphics[width=2in]{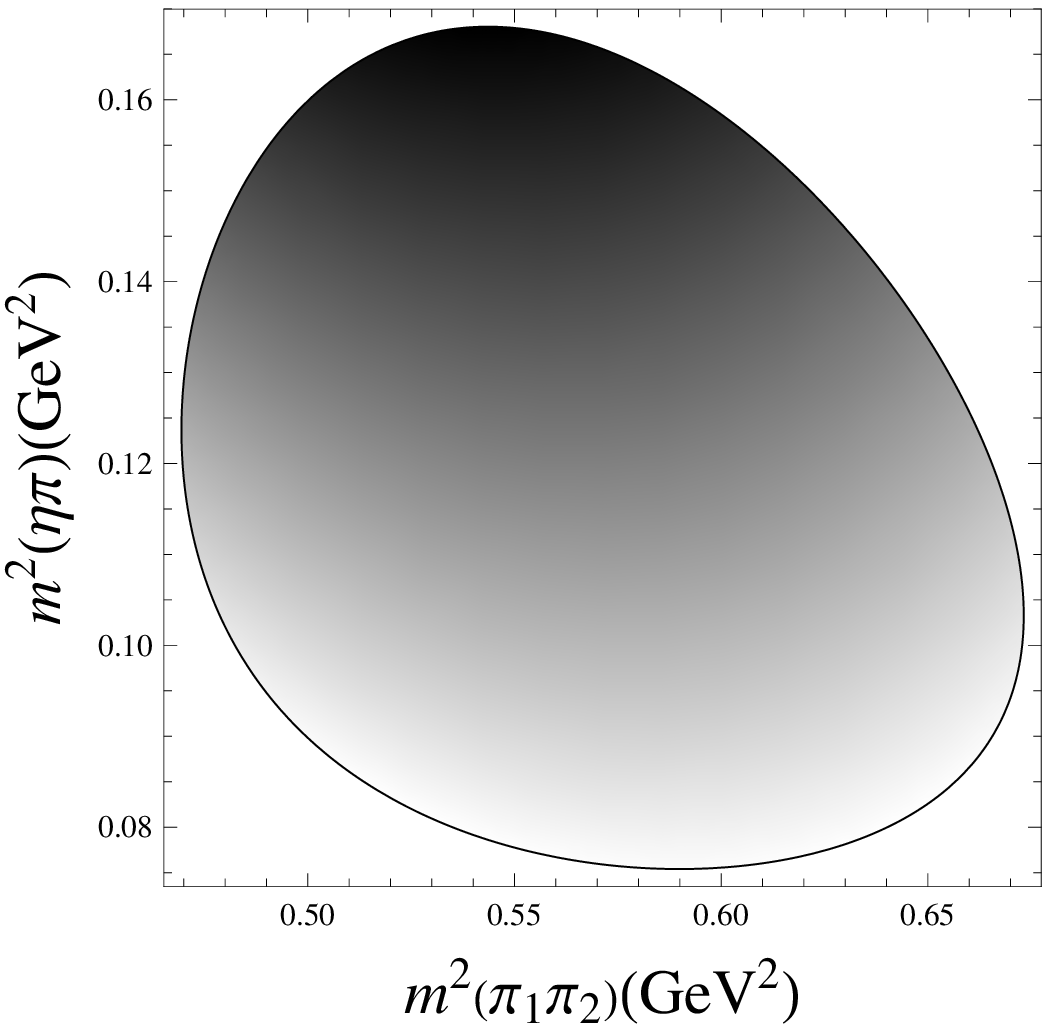}
  \includegraphics[width=2in]{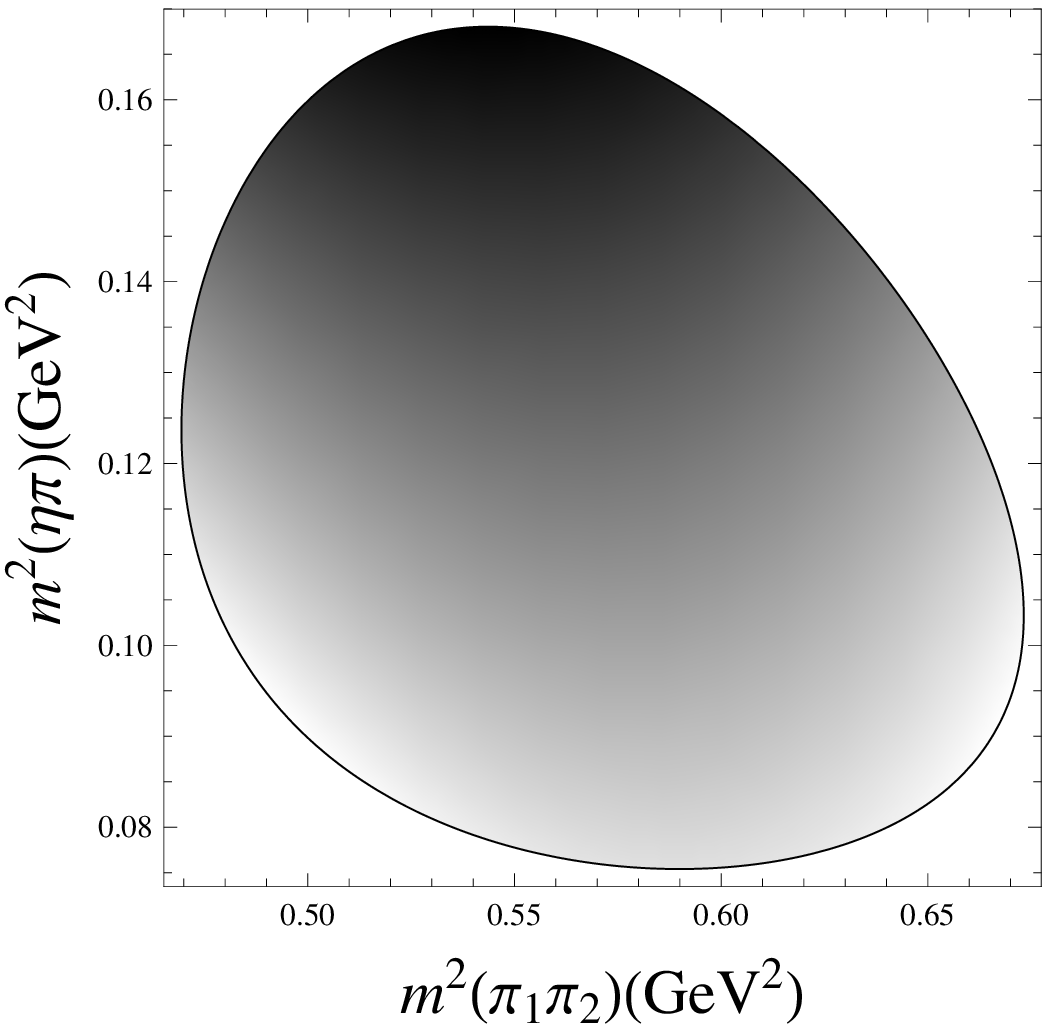}
  \caption{Dalitz plot prediction from Large-$N_c$ ChPT (left) and from Large-$N_c$ RChT (right)}\label{Dalitz}
\end{figure}

In addition to the branching ratio, the study of the shape of the differential decay width (Fig.~\ref{LNRspect}) and the Dalitz plot (Fig.~\ref{Dalitz}) provides important information. Rewriting the amplitude in terms of the following kinetic variables

\begin{equation}
x\equiv \Frac{1}{\sqrt{3}}\,\Frac{(T_1-T_2)}{\bra T\ket}\, ,
\qquad \qquad
y\equiv \Frac{1}{3}\left(2+\Frac{m_\eta}{m_\pi}\right)
\,\Frac{T_3}{\bra T\ket} \,-\, 1\, ,
\end{equation}
with $T_1=\frac{u-(m_{\eta'}-m_\pi)^2}{2 m_{\eta'}}$,
$T_2=\frac{t-(m_{\eta'}-m_\pi)^2 }{2 m_{\eta'}}$,
$T_3=\frac{s-(m_{\eta'}-m_\eta)^2 }{2 m_{\eta'}}$
and $\bra T\ket=\frac{1}{3}(T_1+T_2+T_3)=\frac{1}{3}(2 m_\pi+m_\eta-m_{\eta'})$, one can expand in terms of $x$ and $y$ and compare with the experimental description \cite{Borasoy}:
\vspace{-0.5cm}
\begin{equation}\label{oldparam}
|\mM|^2=|N|^2[1+ ay + by^2 + cx^2].
\end{equation}
In our approach, which preserves $C$-parity, the odd powers of $x$ are absent. The terms $ay$ and $cx^2$ are dominant: $a=-0.42$ and $c=-0.12$. However, the contribution $by^2$ is found to be of the same order as other subleading terms in the $x^2$ expansion. Thus, we propose a more suitable functional form for the Dalitz plot analysis, also including the subdominant terms of similar magnitude:
\begin{equation}\label{newparam}
|\mM|^2=|N|^2[1+ (ay + cx^2) + (by^2 + \kappa_{21}x^2y+\kappa_{40}x^4)].
\end{equation}
We obtain $b=14\cdot10^{-3}$, $\kappa_{21}=26\cdot10^{-3}$ and $\kappa_{40}=4\cdot10^{-3}$. This two extra terms should be included in future experimental analysis. Otherwise, the value obtained for $b$ will result completely distorted as it will try to reproduce the absent terms. We also find interesting relations among these coefficients that only depend, at the considered order, on precise combinations of pseudoscalar masses: $a/c=3.40$ and $\kappa_{40}/\kappa_{21}=0.15$ and the inequality $b/a<0$.

\section{Large-$N_c$ RChT}

Resonance Chiral Theory (RChT) \cite{RChTa} is the most general chiral invariant theory including the Goldstones Bosons from the spontaneous chiral symmetry breaking and the mesonic resonances. At Large-$N_c$ the amplitude for the process $\eta'\rightarrow\eta\pi\pi$ (parameterized as in the previous section) has two OZI-suppressed components (at the order considered), $\mM_{\eta_s\eta_q\pi\pi}=\mM_{\eta_s\eta_s\pi\pi}=0$, leaving only $\mM_{\eta_q\eta_q\pi\pi}$\footnote{In the low energy limit, the Large-$N_c$ result is recovered using the relations $L_8=c_m^2/2M_S^2$, $L_5=c_dc_m/M_S^2$ and $3L_2+L_3=c_d^2/2M_S^2$ \cite{RChTa}.}:

\begin{eqnarray}
&&\mM_{\eta'\to\eta\pi^+\pi^-} =
c_{qq}\,\,\times\,\, \Frac{1}{  F_\pi^2} \,
\left[  \, \Frac{m_\pi^2}{2} \,\,\, +\,\, \,
\Frac{4 c_d c_m}{F^2}\, \Frac{m_\pi^4}{M_S^2}
\right.  \\
&+&\Frac{1}{F^2}\Frac{\left[ c_d(t-m_\eta^2-m_\pi^2)+2 c_m^2 m_\pi^2\right]
\,\left[c_d (t-m_{\eta'}^2-m_\pi^2)+2 c_m m_\pi^2\right]}{M_{a_0}^2-t}\nonumber\\
&+&\Frac{1}{F^2}  \Frac{\left[ c_d (u-m_\eta^2-m_\pi^2)+2c_m^2 m_\pi^2\right]
\,\left[c_d  (u-m_{\eta'}^2-m_\pi^2)+2c_m m_\pi^2\right]}{M_{a_0}^2-u}\nonumber\\
&+&\left.\, \Frac{1}{F^2} \left[ c_d (s-m_\eta^2-m_{\eta'}^2)+2c_m^2 m_\pi^2\right]
\,\left[c_d (s-2 m_\pi^2)+2c_m m_\pi^2\right]
\, \times \,
\left\{ \Frac{\cos^2\phi_S}{M_{\sigma}^2-s}
+\Frac{\sin^2\phi_S }{M_{f_0}^2-s}\right\}\, \right]
\, ,  \nonumber
\end{eqnarray}

\noindent
with $F_{\pi}=F\big(1+\frac{4c_dc_m}{F^2}\frac{m_\pi^2}{M_S^2}+{\cal O}(m_\pi^4)\big)$ and the scalar multiplet is taken at $M_S=980$ MeV. $c_m$ is less important since always enters in the amplitude multiplied by $m_{\pi}^2$ and we choose $c_m=F^2/4 c_d$ \cite{cdcmrelation}.  However, for $c_d$ one finds several values in the literature and finally we choose $c_d=(26\pm 9)$ MeV \cite{Guo-aIJ} (see Ref.~\cite{RJJP} for the discussion). On the other hand, since the decay rate is essentially proportional to $c_d^4$, using the experimental value $\mathfrak{B}(\eta'\rightarrow\eta\pi^+\pi^-)=(44.6\pm1.4)\%$ one obtains a precise theoretical prediction for this parameter, $c_d=(28.9\pm0.2)$ MeV. Likewise, in this approach the decay width is found to be dominated by the $a_0$ exchanges, being the contribution from the isoscalar scalar far too small.

In the study of the shape of the differential decay width, using the parametrization proposed in Eq.~(\ref{newparam}), we obtain $a=-0.13$, $c=-0.07$, $b=0$, $\kappa_{21}=-6.6\cdot10^{-3}$, and $\kappa_{40}=-1.3\cdot10^{-3}$. The prediction for $a$ and $c$ agrees with the VES experimental analysis\footnote{Our parameter $c$ defined in Eq. \ref{newparam} corresponds to the parameter $d$ in Ref.~\cite{ExpDalitz}.} \cite{ExpDalitz}. However, the prediction for $b$ disagrees. This fact can be understood taking into account that in Ref.~\cite{ExpDalitz} $\kappa_{21}$ and $\kappa_{40}$ were not used and then the $b_{exp}$ could be distorted. Similar relations found among these parameters in the previous section are also found here but with slightly different numbers: $a/c=2.0$ and $\kappa_{40}/\kappa_{21}=0.2$. Likewise, one also finds $b/a<0$.





\end{document}